\documentstyle[aps,prb,preprint,psfig]{revtex}

\begin{document}
\draft
\title{$\Psi=0$ at a Sharp Semiconductor/Insulator Interface. 
Is~This~Correct?}            

\author{Leonid S.~Braginsky\thanks{Electronic address: 
brag@isp.nsc.ru}}
\address{Institute of Semiconductor Physics, 630090, 
Novosibirsk, Russia}
\date{\today}
\maketitle
\begin{abstract}
The generalized boundary conditions for the envelope wave 
function that take into account the real structure of an 
interface were used to investigate the hole spectrum of 
the semiconductor quantum dot embedded in an insulator matrix.
An essential influence of the interface levels, which could 
exist at the top of the valence band, on the hole spectrum has 
been demonstrated.  It is found that usually applied boundary 
conditions, wherein all components of the envelope wave function 
vanish at the interface, can be used only 
in the absence of the interface levels close to the band edge.  
\end{abstract}
\narrowtext
To determine the electron states in semiconductor quantum dots, 
the envelope function approximation is  applied.\cite{Woggon} 
The envelope wave functions are usually supposed to be vanished 
at the interface in order to describe the quantum confinement. 
If the intervalley or interband  degeneracy of the electron 
spectrum  occurs in the semiconductor, then each component of 
the envelope wave function  is assumed to be vanished at the 
interface.\cite{Woggon,Xia,Ekimov}  The last statement, 
however, needs to be justified.

Indeed, the boundary condition $\psi=0$ for the proper (not 
envelope) wave function arises from the solution of 
the Schr\"odinger equation for the infinitely high  step-like 
potential barrier.  Such a barrier can't be considered in the 
framework of the envelope function approximation wherein the 
potential must be smooth on the scale of the lattice constant. 
This means that  the boundary condition $\Psi=0$ is justified if 
the potential that restricts the electron movement is smooth on 
the scale of the  lattice constant, but  sharp on the scale 
of the electron wavelength. Perhaps, this happens at the 
contacts of chemically similar materials (e.g., 
GaAs/AlAs),\cite{Volkov1} but not at the contacts of too 
different materials (e.g., semiconductor microcrystals embedded 
in the glass matrix). 

The interface influence on the electrons in the quantum dot has 
to be more complicated if the simple  boundary condition 
$\Psi=0$ doesn't hold. It is well known that  band mixing 
exists  at the interface in the heterojunctions of A$_{\rm 
III}$B$_{\rm V}$ semiconductors.\cite{Ivchenko,Foreman} 
However, this is impossible at the plane interface where 
$\Psi=0$. The bound electron states are also possible at the 
interface owing to the interband mixing;\cite{Volkov} in these 
states the electron wave function behaves roughly as 
$\psi\propto [\exp{(-\gamma_1 r)}-\exp{(-\gamma_2 r})]$, i.e., 
the wave function vanishes at the interface and far from it, but 
has an extremum at a certain distance from the interface.

The band mixing arises when the crystal symmetry that leads 
to the interband degeneracy in the bulk disappears at the 
interface.  The lattice constant is the size that is 
characteristic for such symmetry. Therefore the proper 
boundary conditions for the envelope wave functions should take 
into account the real structure of the interface. 

In this paper we propose simplest boundary conditions
that take into account these subtle details of 
the interface influence. We find the conditions under that the 
boundary condition $\Psi=0$ is applicable at a sharp 
interface.

Let $z=0$ be the plane interface between a semiconductor ($z>0$) 
and an insulator ($z<0$). Assume the two-fold degeneracy for the 
electron band  of the semiconductor and nondegenerate 
electron band in the insulator.  Then
the boundary conditions for the 
envelope wave functions of the electrons in these bands can be 
written as follows:\cite{PR98}
\begin{eqnarray}
\label{2_1}
&&\Psi_1(\tau_1^0)=b_{11}\Psi_l(\tau_{11}), \nonumber \\
&&\Psi_2(\tau_2^0)=b_{22}\Psi_l(\tau_{22}), \\
&&b_{31}\Psi_1(\tau_{31})+b_{32}\Psi_2(\tau_{32})    
=\Psi_l(\tau_3^0).\nonumber 
\end{eqnarray}
Where $\Psi_1$ and $\Psi_2$ are the envelopes that relevant
to the degenerate band of the semiconductor, and $\Psi_l$ is the 
electron envelope in the insulator. To obtain the parameters  
$b_{ij}$ and $\tau_{ij}$ of the boundary conditions (\ref{2_1}), 
the Shr\"odinger equation has to be solved in the narrow (about 
a few lattice constants) region at the interface.  It is 
impossible at an arbitrary and rather imperfect interface.  
Nevertheless, these parameters  are independent of the electron 
energy; they characterize the interface, and estimations of 
their values (the small width of the  interface region is the 
fact that is important for  these estimations) are $b_{ij}\sim 
1$ and $|\tau_{ij}|\sim a$, where $a$ is the lattice constant.  
Thus, the boundary conditions (\ref{2_1}) take into account the 
real structure of the interface. 

We assume the effective-mass approximation holds in the bulk of 
each material, so that $\Psi(\tau)=\Psi(0)+\tau\Psi'(0)$.  The 
large bands offset at the interface restricts the electron 
movement.  If so, then $\Psi_l\propto\exp{(\gamma_l z)}$ and 
$\Psi'_l=\gamma_l\Psi_l$, where the $\gamma_l$ value can be 
considered as independent of the electron energy. Eliminating 
$\Psi_l$ from the Eqs.~(\ref{2_1}) yields 
\begin{eqnarray} 
\label{2_2} 
&&\Psi_1(\tilde{\tau}_{11})+\tilde{b}_{12}\Psi_2(\tilde{\tau}_{12})=0,\\
&&\tilde{b}_{21}\Psi_1(\tilde{\tau}_{21})+\Psi_2(\tilde{\tau}_{22})=0,
\nonumber 
\end{eqnarray}
where  $\tilde{b}_{ij}\sim b_{ij}$ and $\tilde{\tau}_{ij}\sim 
\tau_{ij}$ are known functions of $b_{ij}$, $\tau_{ij}$, and 
$\gamma_l$.
To ensure the probability flux conservation at the interface,
we have to assume 
\begin{equation}
\label{2_3}
\frac{\tilde{b}_{12}(\tilde{\tau}_{22}-\tilde{\tau}_{12})}{m_1}=
\frac{\tilde{b}_{21}(\tilde{\tau}_{11}-\tilde{\tau}_{21})}{m_2},
\end{equation}
where $m_1$ and $m_2$ are  effective masses of the 
appropriate bands. 

The Eqs.~(\ref{2_2}) are  the general form of the boundary 
conditions that  should be written instead of $\Psi=0$ at a 
sharp semiconductor/insulator  interface.  The most general 
boundary conditions that are applicable at such interface have 
been considered in Ref.~\cite{Volkov}. Ours, Eqs.~(\ref{2_2}), 
hold in the effective-mass approximation.  This approximation 
has been used in \cite{PR98} to obtain the Eqs.(\ref{2_1}) and 
to estimate the parameters $b_{ij}$ and $\tau_{ij}$.

It is important that the boundary conditions (\ref{2_2}) are 
nonlocal; they relate the envelopes at the different points 
$\tilde{\tau}_{ij}$ near the interface.   However, the mean 
width of the ''nonlocality region'' is small in comparison with 
the electron wavelength $\lambda$ ($|\tilde{\tau}_{ij}| \sim 
a\ll \lambda$) .  To understand consequences of this 
nonlocality, let, at first, assume $\tilde{\tau}_{ij}=0$.  Then 
the equations (\ref{2_2}) become homogeneous in $\Psi_{1,2}$, 
and so their nonzero solutions exist only when 
\begin{equation} 
\label{2_4}
1-\tilde{b}_{21}\tilde{b}_{12}=0.
\end{equation}

To be precise, for the parameters $\tilde{b}_{ij}$ that
do not obey the Eq.~(\ref{2_4}), the envelopes $\Psi_{1,2}(0)$ 
are as small as $\tau\Psi'(0)$, i.e., $\Psi_{1,2}(0)\sim 
a/\lambda \rightarrow 0$; this is the accuracy, under which the 
 simple boundary conditions $\Psi_{1,2}(0)=0$ are applicable. 
 They are not applicable if Eq.~(\ref{2_4}) holds. It can be 
 shown that the condition (\ref{2_4}) means the proximity of a 
certain interface level to the band edge.  The energy position 
of this level is determined by the parameters $\tilde{b}_{ij}$ 
and $\bbox{\tilde{\tau}}_{ij}$, i.e., by  structure of the 
interface.

Thus, the simple boundary conditions 
$\Psi_{1,2}(0)=0$ can be used at a sharp interface in the 
absence of interface levels close to the band edge. Otherwise, 
the  general boundary conditions (\ref{2_2}) should be used. 

It should be noted that assumption of the large bands offset at 
the interface is not important for our  consideration. 
The boundary conditions (\ref{2_1}) could be used in that 
case.  This means that  the 
simple boundary conditions $\Psi_{1,2}=0$ can be used  at a 
sharp interface even in the absence of real potential barrier 
there, provided that the interface levels are not close to 
the band edge. In that case the quantum confinement arises  
because the resonant tunneling of  electrons is no longer 
possible through the interface.\cite{Zhu}

It is possible to rewrite the Eqs.~(\ref{2_2}) in the more 
simple form:  
\begin{equation} 
\label{2_5} 
\left( 
\begin{array}{l} 
\Psi_1 \\ \Psi'_1
\end{array}
\right)=
\left(
\begin{array}{ll}
t_{11}&t_{12}\\
t_{21}&t_{22}
\end{array}
\right)
\left(
\begin{array}{l}
\Psi_2 \\ \Psi'_2
\end{array}
\right),
\end{equation}
where  
\begin{eqnarray*}
&&\begin{array}{ll}
t_{11}=\displaystyle\frac{\tilde{b}_{12}\tilde{b}_{21}
\tilde{\tau}_{21}-\tilde{\tau}_{11}}
{\tilde{b}_{21}(\tilde{\tau}_{11}-\tilde{\tau}_{21})}, &
t_{12}=\displaystyle 
\frac{\tilde{b}_{12}\tilde{b}_{21}\tilde{\tau}_{12}
\tilde{\tau}_{21}-\tilde{\tau}_{11}\tilde{\tau}_{22}}
{\tilde{b}_{21}(\tilde{\tau}_{11}-\tilde{\tau}_{21})}, 
\\   \ &\ \\
t_{21}=\displaystyle\frac{1-\tilde{b}_{12}\tilde{b}_{21}}
{\tilde{b}_{21}(\tilde{\tau}_{11}-\tilde{\tau}_{21})}, &
t_{22}=\displaystyle-\frac{\tilde{b}_{12}\tilde{b}_{21}
\tilde{\tau}_{12}-\tilde{\tau}_{22}}
{\tilde{b}_{21}(\tilde{\tau}_{11}-\tilde{\tau}_{21})}.
\end{array}
\end{eqnarray*}
Then the Eq.(\ref{2_4}) takes the form 
$t_{21}=0$.

The interface influence on the electrons is determined by the
parameters $t_{ij}$. They are not independent. It follows from 
the Eq.~(\ref{2_3}) that the determinant of the $||t_{ij}||$ 
matrix is equal to $m_1/m_2$. Moreover, $t_{12}\sim a$, and  so  
it is possible to assume  $t_{12}=0$ by the appropriate choice 
of the position of the  plane $z=0$ within the unit cell at the 
interface.

Thus, there are two parameters, $t_{11}\sim 1$ and   $t_{21}\sim 
a^{-1}$, that determine the interface influence on the electron.
One of them, $t_{21}$, is sensitive to the position of the 
interface level: it  vanishes when this level coincides with 
the band edge. Another one, $t_{11}$, can be considered as a 
trial parameter.

To consider the hole states in the spherical quantum dot, we 
write the Luttinger Hamiltonian in the spherical approximation 
as follows:\cite{Luttinger}
\begin{equation}
\label{1_1}
\hat{H}=\left(\gamma_1+\frac{5}{2}\gamma\right)\frac{\hat{p}^2}{2m_0}
-\frac{\gamma}{m_0}\left(\hat{\bf p}{\bf J}\right)^2,
\end{equation}
where $\hat{\bf p}$ is the momentum and ${\bf J}$ are the 
$4\times 4$  matrices of the angular moment J=3/2; $\gamma>0$ 
and $\gamma_1$ are the Luttinger parameters that relevant to the 
light and heavy effective masses of the holes:  
$m_l=m_0(\gamma_1+2\gamma)^{-1}$ and 
$m_h=m_0(\gamma_1-2\gamma)^{-1}$, $m_0$ is  mass of the free 
electron. 

The moment $F=1/2,\ 3/2, ...,$ and its projection $M$ are the 
good quantum numbers due to the spherical symmetry.
Solutions of the Schr\"odinger equation with the 
Hamiltonian~(\ref{1_1}) are of the form \cite{Gel'mont}
\begin{eqnarray*}
&&\psi_{EM}(r,\theta,\varphi)=\sqrt{2F+1}
\sum_l(-1)^{l-3/2+M}R_{Fl}(r)\\
&&\phantom{\psi_{EM}(r,\theta,\varphi)}\times
\sum_{m\mu}\left(
\begin{array}{lll}
l&3/2&F\\
m&\mu&-M
\end{array}
\right)
Y_{lm}(\theta,\varphi)\chi_\mu.
\end{eqnarray*}
Where $\left(
\begin{array}{lll}
l&3/2&F\\
m&\mu&-M
\end{array}
\right)
$
are the Wigner symbols, and $\chi_\mu$ is the eigenvector of 
the $J_z$ matrix.
The radial functions $R_{F,F+1/2}$ and  $R_{F,F-3/2}$ that 
relevant to the even solutions obey the equations  
\cite{Gel'mont} 
\widetext
\begin{eqnarray}
\label{1_2}
&&(\gamma_1-2\gamma\cos\alpha_F)P^+_FP_FR_{F,F+1/2}
+2\gamma\sin\alpha_FP^+_FP_{-F}R_{F,F-3/2}+\frac{2m_0}{\hbar^2}
[E-U(r)] R_{F,F+1/2}=0,
\nonumber \\
&&(\gamma_1+2\gamma\cos\alpha_F)P^+_{-F}P_{-F}R_{F,F-3/2}
+2\gamma\sin\alpha_FP^+_{-F}P_FR_{F,F+1/2}+\frac{2m_0}{\hbar^2}
[E-U(r)] R_{F,F-3/2}=0,\\
&&\mbox{where }\;\;
\cos\alpha_F=\frac{2F-3}{4F},\;\;\;\sin\alpha_F\geq 0,\;\;\;
P_F=\frac{d}{dr}+\frac{F+3/2}{r},\;\;\;
P^+_F=\frac{d}{dr}-\frac{F-1/2}{r}.
\nonumber
\end{eqnarray}
\narrowtext

Let us, at first, suppose that Eqs.~(\ref{1_2}) hold also 
at the interface where the potential $U(r)$ restricts the 
hole movement. Then we can obtain the boundary conditions for 
the radial wave functions.  Two of them 
arise after  integration of the Eqs.~(\ref{1_2}) over the narrow 
region $|r-r_0|<w/2$ $(a\ll w\ll \lambda)$ at the interface.  To 
obtain 
another two boundary conditions, we have to multiply the 
Eqs.~(\ref{1_2}) by $r-r_0$ before the integration. After 
elimination of $R_{F,F+1/2}(r_0+w/2)\propto 
\exp(-\gamma_{F+1/2} r)$ and $R_{F,F-3/2}(r_0+w/2) \propto 
\exp(-\gamma_{F-3/2} r)$ (where $\gamma_{F+1/2}>0$ and 
$\gamma_{F-3/2}>0$ are the decay exponents of the wave functions 
off the dot boundary)
from the derived equations, we obtain
\begin{eqnarray}
\label{1_3}
&&-\gamma_{F+1/2}R_{F,F+1/2}+R_{F,F+1/2}'=
\frac{(\gamma_{F+1/2}W_++V_+)(\gamma_1+2\gamma\cos 
\alpha_F) -2\gamma\sin \lambda_F(\gamma_{F+1/2}W_-+V_-)}
{\gamma_1^2-4\gamma^2},\\
&&-\gamma_{F-3/2}R_{F,F-3/2}+R_{F,F-3/2}'=
\frac{(\gamma_{F-3/2}W_-+V_-)(\gamma_1-2\gamma\cos 
\alpha_F) -2\gamma\sin 
\lambda_F(\gamma_{F-3/2}W_++V_+)}{\gamma_1^2-4\gamma^2}. 
\nonumber 
\end{eqnarray}
Where $V_+=2m_0\int_{-w/2}^{w/2}U(r-r_0)R_{F,F+1/2}\,dr$,
$V_-=2m_0\int_{-w/2}^{w/2}U(r-r_0)R_{F,F-3/2}\,dr$,
$W_+=-2m_0\int_{-w/2}^{w/2}(r-r_0)U(r-r_0)R_{F,F+1/2}\,dr$, and
$W_-=-2m_0\int_{-w/2}^{w/2}(r-r_0)U(r-r_0)R_{F,F-3/2}\,dr$.
These values  vanish when $w\rightarrow 0$, if  the 
potential $U(r)$  has not any singularity at the 
interface.  This leads to the simple boundary conditions 
$R_{F,F+1/2}(-\hbar/\gamma_{F+1/2})
=R_{F,F-3/2}(-\hbar/\gamma_{F-3/2})=0$.

The values of $V_{\pm}$ and $W_{\pm}$ don't vanish at a 
sharp interface where the potential $U(r)$ changes essentially   
on the scale of the lattice constant. In particular, this is 
possible at a strain interface due to mismatch of the lattice 
constants of the bordering materials (e.g., at the Ge/Si 
interface).  Then $U(r)$ can be estimated as $U\sim D(\delta 
a/a)$ for $|r-r_0|<a$, where $D\sim 10\,$eV is the constant of 
the deformational potential and $\delta a$ is the lattice 
mismatch.  So that $W\sim \delta a/a \sim 1$ and $V\sim \delta 
a/a^2$. The boundary conditions (\ref{1_3}) accept the form of 
the Eqs.~(\ref{2_2}) after expansion of the radial wave 
functions in the integrands. This is the case even when 
$\gamma_{F-3/2}\rightarrow \infty$ and 
$\gamma_{F+1/2}\rightarrow \infty$.

We shall use the boundary conditions (\ref{2_2}) in the form 
(\ref{2_5}) to obtain the hole spectrum of the quantum dot. The 
radial wave functions in the free space $(U=0)$ are\cite{Ekimov}
\begin{eqnarray}
\label{3_1}
&&R_{F,F+1/2}(r)=Aj_{F+1/2}(kr)+Bj_{F+1/2}(kr\sqrt{\beta}),\\
&&R_{F,F-3/2}(r)=A_1j_{F-3/2}(kr)+B_1j_{F-3/2}(kr\sqrt{\beta}),
\nonumber 
\end{eqnarray}
where $j_l(z)$ are the spherical Bessel functions, 
$A_1=A\tan(\alpha_F/2)$, $B_1=B\cot(\alpha_F/2)$, 
$\cos\alpha_F=(2F-3)/(4F)$, $\sin\alpha_F\geq 0$, 
$\beta=m_l/m_h$; $A$ and $B$ are the constants that are 
determined by the boundary conditions at $r=r_0$.  

By substitution (\ref{3_1}) into the boundary conditions 
(\ref{2_5}) we obtain the system of equations which 
is homogeneous in $A$ and $B$.
Its nonzero solutions exist only when the determinant  
vanishes, i.e.,
\widetext
\begin{eqnarray}
\label{3_2}
&&\left[t_{11}j_{F+1/2}(kr_0)-\tan\frac{\alpha_F}{2}
j_{F-3/2}(kr_0)\right]
\left[t_{21}j_{F+1/2}(kr_0\sqrt{\beta})+t_{22}j'_{F+1/2}(kr_0\sqrt{\beta})
+\cot\frac{\alpha_F}{2}j'_{F-3/2}(kr_0\sqrt{\beta})\right]\nonumber \\
&&-\left[t_{21}j_{F+1/2}(kr_0)+t_{22}j'_{F+1/2}(kr_0)
-\tan\frac{\alpha_F}{2}j'_{F-3/2}(kr_0)\right]
\left[t_{11}j_{F+1/2}(kr_0\sqrt{\beta})+\cot\frac{\alpha_F}{2}
j_{F-3/2}(kr_0\sqrt{\beta})\right]=0.
\end{eqnarray}
Where $j'\equiv dj/dr$.   The
Eq.~(\ref{3_2}) determines the hole spectrum of the quantum dot: 
$E_n=(\gamma_1-2\gamma)\hbar^2k_n^2/2m_0$, where $k_n$ are the
roots of Eq.~(\ref{3_2}).  Influence of the interface on this 
spectrum is determined by the parameters $t_{ij}$.  To estimate 
the energy $E_0$ of the interface hole state, we assume 
$k=i\kappa$, where $\kappa r_0\gg 1$. Then from 
the Eq.~(\ref{3_2}) we obtain

\begin{equation}
\label{3_3}
\kappa\simeq\frac{t_{21}\left(\tan\frac{\alpha_F}{2}
+\cot\frac{\alpha_F}{2}\right)}
{(t_{11}t_{22}-1)(1-\sqrt{\beta})+
t_{11}\left(\sqrt{\beta}\cot\frac{\alpha_F}{2}
+ \tan\frac{\alpha_F}{2}\right)
-t_{22} \left(\sqrt{\beta}\tan\frac{\alpha_F}{2}
+ \cot\frac{\alpha_F}{2}\right)}.
\end{equation}
\narrowtext
So that $E_0=-\hbar^2\kappa^2/2m_h$.
The simple case that 
corresponds to $\Psi_{1,2}(0)=0$ follows from the 
Eq.~(\ref{3_2}) if we assume there $t_{21}\rightarrow \infty$.  
This is possible when $t_{21}\gg k$.  The value of $t_{21}$ can 
be estimated from Eq.~(\ref{3_3}), $t_{21}\sim 
\kappa=\hbar^{-1}\sqrt{2m_h|E_0|}$.  Therefore the boundary 
conditions  $\Psi_{1,2}(0)=0$ are applicable at a sharp 
interface, if $|E_0|\gg\hbar^2k^2/2m_h$, i.e., when the energy 
of the interface level much exceeds the energy of the hole. 
Otherwise, the general boundary conditions (\ref{2_5}) should be 
used.

Figure 1 displays the left side of the Eq.~(\ref{3_2}) as a 
function of $kr_0$. We assume $t_{11}=1$, $m_h=m_0$, $\beta=0.1$ 
and obtain $t_{21}$ from the Eq.~(\ref{3_2}) provided 
$E_0=0.01\,$eV.  The dashed curve presents the similar 
dependence that follows from the simple boundary conditions 
$\Psi_{1,2}(0)=0$.\cite{Ekimov}  We reveal an essential 
difference between the hole spectra.  Apart from an 
essential change of the position  of the roots of the
Eq.~(\ref{3_2}), we find that some of them become complex 
($kr_0=15.5\pm 1.3i$ and $kr_0=25.3\pm 0.7i$ on Fig.~1), and so 
the relevant hole states become quasistationary and bounded at 
the interface. This could be essential for the optical 
properties of the quantum dot. Moreover, such states affect the 
electron transport in the array of the quantum dots; they 
increase the effective cross section of the quantum dot. Note 
that the solid curve becomes close to the dashed one when $E_0$ 
is about a few eV.

The hole spectrum was found to be sensitive to the 
energy position of the interface level; namely, whether or not 
it is close to the band edge.  Such levels really exist at the 
top of the valence band in some semiconductor/insulator 
contacts;\cite{Fujitani} they are responsible for the 
Fermi-level pinning.  It seems  that the electron interface 
level should be close to the valence band at least  in  wide-gap 
semiconductors.  The interface level becomes empty then it is 
shifted too far off the top of the valence band.  This results 
in a large surface charge and a strong band bending that is not 
favorable from the energetical point of view.  Nevertheless, the 
interface level can be shifted as the result of the structure 
reconstruction of the interface. Such reconstruction does not 
essentially affect the interatomic spaces or angles, but it 
makes the interface level to be closer to the top of the valence 
band.

In conclusion, we propose the general boundary conditions 
for the envelope wave functions to investigate the hole spectrum 
of the spherical quantum dot. We show that usually applicable 
boundary conditions $\Psi_{1,2}(0)=0$ can be used at a smooth 
interface or at a sharp one provided that the energy 
separation of nearest to the band edge interface level much 
exceeds the energy of the hole under consideration.  Two real 
parameters are sufficient to determine an interface influence  
on the hole spectrum. They could be measured in optical 
experiments or estimated theoretically [e.g., from 
Eq.~(\ref{1_3})] for a certain model of the interface structure.  
The boundary conditions (\ref{2_5}) can be used also to describe 
the intervalley mixing of the electron in the conduction band.

\acknowledgments
I  wishe to thank  
Prof.\ V.\ A.\ Volkov  for helpful discussions, and Prof.\ B.\ 
A.\ Foreman for reprint,\cite{Foreman} which was sent to me 
before the publication.  This work was supported by the Russian 
Foundation for the Basic Research, Grant No.~99-02-17019.

\begin{figure}
\centerline{\psfig{figure=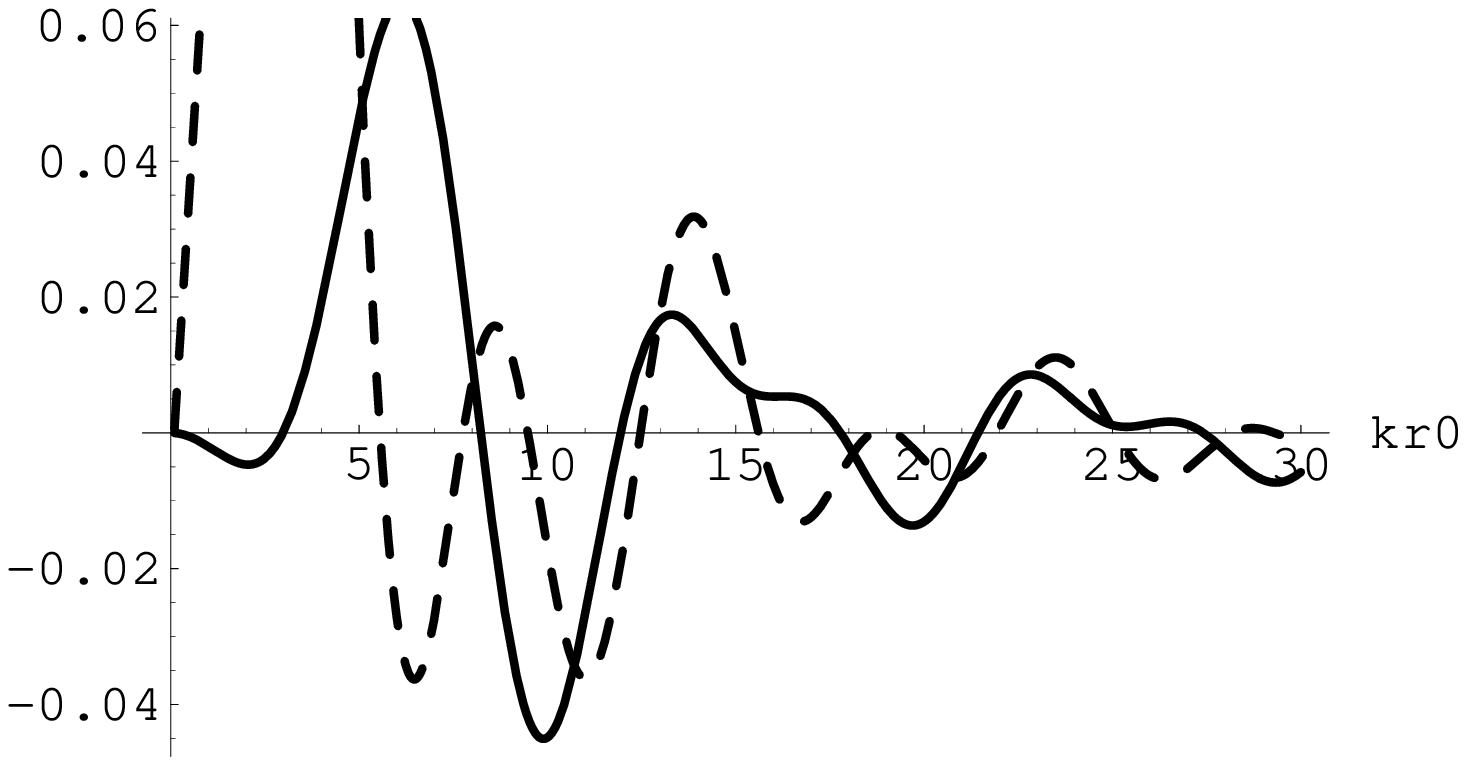,height=4.0in}}
\caption{Left side of Eq.~\protect(\ref{3_2}) as a function 
of $kr_0$ (bold curve). Similar dependence which relevant to 
the boundary conditions $\Psi_{1,2}(0)=0$  (dashed curve).}

\label{fig1}
\end{figure}

\end{document}